\documentclass{ws-ijmpb}
\usepackage{tikz}

\begin{document}


%
%

\title{Fluctuation theorems without time-reversal symmetry}

\author{Chenjie Wang}

\address{James Franck Institute, Department of Physics, University of Chicago\\
Chicago, Illinois 60637, USA\\
cjwang@uchicago.edu}

\author{D. E. Feldman}

\address{Department of Physics, Brown University\\
Providence, Rhode Island 02912, USA\\
dmitri\_feldman@brown.edu }

\maketitle


\begin{abstract}

Fluctuation theorems establish deep relations between observables away from thermal equilibrium. Until recently, the research on fluctuation theorems was focused on time-reversal-invariant systems. In this review we address some newly discovered fluctuation relations that hold without the time-reversal symmetry, in particular, in the presence of an external magnetic field. One family of relations connects non-linear transport coefficients in the opposite magnetic fields. Another family relates currents and noises at a fixed direction of the magnetic field in chiral systems, such as the edges of some quantum Hall liquids. We  review the recent experimental and theoretical research, including the controversy about the microreversibility without the time-reversal symmetry, consider the applications of fluctuation theorems to the physics of topological states of matter, and discuss open problems.

\end{abstract}


\section{Introduction}

From the very beginning of condensed matter physics, solid state theory relied on two complementary approaches: model-building and the use of general principles. The atomistic model was invoked by early crystallographers to explain the laws of crystal habit \cite{hauy}; more than 400 years ago, Kepler speculated about the connection of the shape of a snowflake with the dense packing of spheres \cite{kepler}. The simple Drude \cite{drude} and Sommerfeld \cite{sommerfeld} models of metals were followed by a more realistic band theory of solids \cite{bloch,kane-hasan}. Later, improved models that incorporate electron interactions led to some of the greatest triumphs of condensed matter
physics including the Fermi-liquid theory \cite{Gabriele} and the explanation of superconductivity \cite{BCS-book}. Any recent issue of this journal contains articles on the Hubbard, Tomonaga-Luttinger, Kondo or other models of materials.

In some cases it is  possible to obtain nontrivial predictions without the use of models, solely from the basic principles of quantum and statistical mechanics. Thermodynamics is a particularly powerful source of such predictions and Einstein famously said \cite{einstein-quote}: ``A theory is the more impressive the greater the simplicity of its premises is, the more different kinds of things it relates, and the more extended is its area of applicability. Therefore the deep impression which classical thermodynamics made upon me.'' Einstein himself made some of the most brilliant predictions from general principles by his masterful use of detailed balance  \cite{einstein1,einstein2}. The fluctuation relations, addressed in this review, are a far-reaching development of his ideas.

Early milestones in the field, opened by Einstein, included the Nyquist formula \cite{johnson,nyquist} and the Onsager reciprocity relations \cite{onsager,casimir}. Their generalization led to the fluctuation-dissipation theorem \cite{FDT} (FDT) and the linear response theory \cite{Kubo}. Another breakthrough in the application of general principles to many-body systems came from the idea of universality  \cite{Kadanoff}. It allowed a precise quantitative description of critical phenomena from  the symmetry considerations without any microscopic  information \cite{ZJ}.

These and other achievements greatly advanced the theory of condensed matter in and close to thermal equilibrium. On the other hand, far from equilibrium, little could be told without the resort to  microscopic models \cite{kynetics}. That situation changed in the 1990s, when the discovery of fluctuation relations brought a powerful general principle to nonequilibrium statistical mechanics \cite{fl-3,fl-2,fl-1,fl0,fl1,fl2}.

The key idea behind fluctuation relations is similar to the principle of detailed balance. Consider a time-reversal-invariant system in an initial state $|\psi_i\rangle$. The system evolves during the time $t_0$. From the  unitarity of quantum mechanics, the probability to find the system in the final state $|\psi_f\rangle$
is exactly the same as the probability to find the system with the initial state $|\psi(t=0)\rangle=\Theta|\psi_f\rangle$ in the final state $|\psi(t=t_0)\rangle=\Theta|\psi_i\rangle$, where $\Theta$ is the time-reversal operator. We now consider a process made of the following three steps:

1) The initial state of the system is measured;

2) The system undergoes unitary evolution over the time interval $t_0$;

3) The final state is determined through measurement.

\noindent
The probability to observe the evolution from $|\psi_i\rangle$ into $|\psi_f\rangle$ is no longer the same as the probability to  observe the evolution from $\Theta|\psi_f\rangle$ to $\Theta|\psi_i\rangle$ since the probabilities to find the system in the initial states
$|\psi_i\rangle$ and $\Theta|\psi_f\rangle$ are not the same away from thermal equilibrium. However, it is often possible to write a simple relation between the two probabilities \cite{crooks}. This, in turn, allows a derivation of numerous relations between  correlation functions of observables. In this review we will be particularly interested in the correlation functions of electric currents.

One might think that all this is of no use in the absence of time-reversal symmetry, e.g., when an external magnetic field is applied. Indeed, most work on fluctuation relations assumes the time-reversal symmetry. However, it transpired recently that a number of nontrivial fluctuation relations hold even without such symmetry. One result \cite{saito} connects systems that transform into each other under the action of the time-reversal operator.  A family of relations has been found
for nonlinear transport coefficients of a system, close to thermal equilibrium, at two opposite directions of the magnetic field. Another result \cite{wang1,wang2}  applies even far away from equilibrium and connects non-linear response and noise at a fixed direction of the magnetic field.
The latter result holds in chiral systems. Since the word ``chirality'' has all too many meanings, we need to explain ours.

By chiral we mean systems, where excitations can propagate in one direction only, that is, either all excitations can propagate only clockwise or all excitations can only propagate counterclockwise. Such transport is known to occur on the edges of  certain quantum Hall liquids and in some other systems. There has been much recent interest in chiral transport in the quantum Hall effect (QHE). The interest comes, in part, from the search for elusive neutral modes \cite{chang}. Besides, the question of chirality proved relevant for the ongoing search for non-Abelian anyons \cite{anyons} in quantum Hall states in the second Landau level (see Section 4). The  more powerful fluctuation relations \cite{wang1,wang2} in chiral systems originate from a stronger form of the causality principle for chiral transport. The standard causality principle states that the past is not affected by the future events. In the chiral case, in addition, one of the following two alternatives holds: either what happens on the right is not affected by the past events on the left or what happens on the left is not affected by the past events on the right.

In this review we address theoretical and experimental work on  fluctuation relations in the absence of the  time-reversal symmetry in chiral and non-chiral systems. Many questions remain open and we discuss future directions in Section 6. We also address the ongoing controversy
about microreversibility  without the time-reversal symmetry.

The paper is organized as follows. In the second section we derive the quantum fluctuation theorem \cite{fl2,tobiska05,andrieux06,esposito07,andrieux09,campisi10} that serves as a foundation for all subsequent discussion. In Section 3, we extract from that theorem the Saito-Utsumi relations \cite{saito} for the  nonlinear transport coefficients of identical systems in the opposite magnetic fields. We also address the verification of the Saito-Utsumi relations in microscopic models and the experimental results \cite{exp1,exp2}. In Section 4 we introduce chiral systems with the emphasis on QHE. We derive fluctuation relations \cite{wang1,wang2} for chiral systems in Section 5. Finally, we address open problems and summarize in Section 6.

\section{Fluctuation theorems}

\begin{figure}[b]
\centering 
\includegraphics{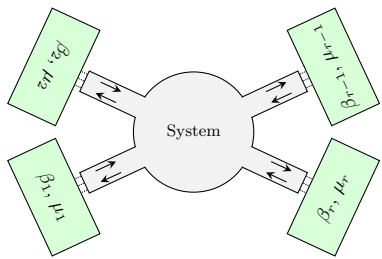}
\caption{Schematics of a quantum open system. The system can exchange energy and particles with $r$ equilibrium reservoirs. Chemical potential difference or temperature difference between any two reservoirs will induce particle current and$/$or heat current.}\label{fig1}
\end{figure}

In statistical mechanics, physical quantities of interest usually undergo random fluctuations. {\it Fluctuation theorems} are a class of exact relations for the distribution functions of those random fluctuations, regardless whether the system under investigation is in an equilibrium or nonequilibrium state. There are fluctuation theorems for, e.g,  entropy in driven isolated systems, work and heat in closed systems, {\it etc}. Fluctuation theorems are typically of the form \cite{fl1,fl2,fl3}
\begin{equation}
 P_F(x) =  P_B(-x) \exp[a(x-b)], \label{ftgeneral}
\end{equation}
where $x$ is the physical quantity or the collection of quantities under investigation, $ P_F(x)$ and $ P_B(x)$ are the distribution functions of $x$ in the so-called {\it forward} and {\it backward} processes respectively, and the constants $a$ and $b$ are determined by the initial conditions of the two processes. The definitions of the forward and backward processes may differ slightly for different systems, but in general they follow these lines: 1) Initially in both processes, the system obeys a Gibbs distribution or can be factorized into several subsystems that obey Gibbs distributions and 2) the dynamical equations (e.g., the Schr\"odinger equation) in the backward process is obtained from the dynamical equations in the forward process by the time-reversal operation. Generally speaking, the system in the forward process and the system in the backward process should not be considered as the same system because they follow different microscopic dynamical equations. However, for a system with the time-reversal symmetry, the dynamical equation is the same in the forward and backward processes. Hence, provided that the initial Gibbs distributions in the two processes are the same,  no difference exists between the two processes for a time-reversal invariant system, and thereby the subindices $F$ and $B$ in (\ref{ftgeneral}) can be dropped.

In this review, we consider systems without the time-reversal symmetry, and focus on a particular fluctuation theorem for energy and particle transport in quantum open systems. This fluctuation theorem is very useful for studying transport phenomena in systems without the time-reversal symmetry, and, particularly, in systems with chirality (see Sections 4 and 5). For a reader, interested in other fluctuation theorems,  many excellent reviews exist, for example, Refs. \refcite{fl1,fl2,fl3}.

The approach of this section builds on Refs. \refcite{andrieux09,campisi10} and closely follows Supplementary information to Ref. \refcite{wang2}.

Let us discuss the specific fluctuation theorem that we are interested in. We consider a setup shown in Fig.~\ref{fig1}: the system in the center is coupled to $r$ reservoirs, each being in equilibrium. The system serves as a bridge, so that energy and particles can be transported between the reservoirs. This setup is commonly used in transport experiments with mesoscopic systems, such as quantum dots, quantum Hall bars, {\it etc}.  A {\it forward} process can then be defined as follows. Initially, at $t\le 0$, the system and reservoirs are decoupled. An interaction $\mathcal V(t)$ that allows particle and energy exchange with the reservoirs is turned on at the times $0\le t\le \tau$. The interaction is turned off at $t\ge \tau$. At $t\le 0$, reservoir $i$ is at equilibrium with the inverse temperature $\beta_i=1/T_i$ and the chemical potential $\mu_i=qV_i$, where $q$ is the charge of a charge carrier and $V_i$ the electric potential. We use only one set of chemical potentials and thus assume that only one carrier type is present which is usually electron. We assume that the size of the system is much smaller than that of the reservoirs, so its initial state is irrelevant in the $\tau\rightarrow \infty$ limit which we will take. It is convenient to regroup the system with one of the reservoirs\cite{andrieux09}, for example, the $r$-th reservoir. The interaction $\mathcal V(t)$ becomes a constant $\mathcal V_0$ when fully turned on during $\tau_0\le t\le \tau-\tau_0$. We assume that $\tau_0\ll\tau $ and $\tau$ is much longer than the relaxation time so that the system remains in a steady state during most of the time interval $\tau$.

We now find the statistical distribution of the changes $\Delta N_i=N_i(t=\tau)-N_i(t=0)$ and $\Delta E_i= E_i(t=\tau) - E_i(t=0)$, where $N_i$ is the particle number and $E_i$ is the energy of the $i$-th reservoir. Let $\mathcal H_i$  and $\mathcal N_i$ be the Hamiltonian and particle number operators of the $i$-th reservoir ($\mathcal H_r$ includes the system). The particle numbers conserve in the absence of $\mathcal V(t)$, i.e., $[\mathcal H_i, \mathcal N_i]=0$. Thus, the initial density matrix factorizes into a product of Gibbs distributions in each reservoir,
\begin{equation}
\rho_{n}=\frac{1}{Z_0^+}\prod_{i} e^{-\beta_i[E_{in}-\mu_i N_{in}]}\label{gibbs}
\end{equation}
where $Z_0^+$ is the initial partition function and the index $n$ labels the quantum state $|\psi_n\rangle$ with the reservoir energies $E_{in}$ and particle numbers $N_{in}$. Here, the ``+'' sign reminds us that we are studying the forward process. An initial joint quantum measurement of  $\mathcal H_i$ and $\mathcal N_i$ is performed at $t=0$, so that the quantum state of the system collapses to a common eigenstate $|\psi_n\rangle$ with the probability $\rho_n$. The state $|\psi_n\rangle$ then evolves according to the evolution operator $ U(t;+)$, which satisfies
\begin{equation}
i\frac{d}{dt} U(t; +) = \mathcal H (t; +) U(t;+), \label{forwardpropagator}
\end{equation}
where the Hamiltonian $\mathcal H(t;+)=\sum_{i} \mathcal H_i +\mathcal V(t)$, and the initial condition is $U(0;+)=1$. At $t=\tau$, a second joint measurement is taken, leading to the collapse of the system to the state $|\psi_m\rangle$ with the reservoir energies $E_{im}$ and particle numbers $N_{im}$.
The probability to observe such process is
\begin{equation}
P[m,n]=|\langle\psi_m|U(\tau;+)|\psi_n\rangle|^2\rho_{n}.
\end{equation}
Hence, repeating the forward process, we obtain the joint distribution function of the energy and particle changes $\Delta E_{i,mn}=E_{im}-E_{in}$ and $\Delta N_{i,mn}=N_{im}-N_{in}$
\begin{align}
P[\Delta\mathbf E, \Delta \mathbf N;+] = &\sum_{mn}\prod_{i=1}^r\delta(\Delta E_i - \Delta E_{i,mn})\delta(\Delta N_i - \Delta N_{i,mn}) \nonumber\\ &\times|\langle\psi_m|U(\tau;+)|\psi_n\rangle|^2\rho_{n} \label{forwarddistri},
\end{align}
where the vectors $\Delta\mathbf E=(\Delta E_1, \dots, \Delta E_r)$ and $\Delta\mathbf N=(\Delta N_1, \dots, N_r)$. Since the total energy and particle number are conserved, one finds that $\sum_i\Delta E_i=\sum_i\Delta N_i=0$. The energy conservation is an approximation due to the time-dependent interaction $\mathcal V(t)$. However, in the limit $\tau\rightarrow \infty$, the violation of the energy conservation is negligible since 
the time-dependence is relevant only  during a short period of time $\tau_0$.

We now study the {\it backward} process.  In that process, the initial temperatures and chemical potentials of the reservoirs are the same as those in the  forward process. Note that such initial temperatures and chemical potentials are not necessarily the same as the final thermodynamic parameters at $t= \tau$ in the forward process for large but finite reservoirs. The time evolution operator $U(t;-)$ is determined by the equation
\begin{equation}
i\frac{d}{dt} U(t; -) = \mathcal H (t; -) U(t;-), \label{backwardpropagator}
\end{equation}
where the Hamiltonian $\mathcal H(t;-)=\Theta \mathcal H(\tau -t;+)\Theta^{-1}$ with $\Theta$ being the time-reversal operator. Here, the ``$-$'' sign stands for the backward process. The $i$-th reservoir has the Hamiltonian $\Theta\mathcal H_i\Theta^{-1}$. Clearly, $\Theta|\psi_m\rangle$ is a common eigenstate of $\Theta\mathcal H_i\Theta^{-1}$ and $\mathcal N_i=\Theta\mathcal N_i\Theta^{-1}$ with the eigenvalues $E_{im}$ and $N_{im}$ respectively.
The amplitude of the transition from the state $\Theta|\psi_m\rangle$ to $\Theta|\psi_n\rangle$ after the time $\tau$ is $(\langle\psi_n|\Theta^{-1}U(\tau;-)\Theta|\psi_m\rangle)^*$.
Hence, performing two quantum measurements at $t=0$ and $t=\tau$ in the backward process, we find that the probability to observe the collapse to the state $\Theta|\phi_m\rangle $ at $t=0$ and the collapse to the state $\Theta |\phi_n\rangle$ at $t=\tau$ is
\begin{equation}
P[m,n]=|\langle\psi_n|\Theta^{-1}U(\tau;-)\Theta|\psi_m\rangle|^2\rho_{m},
\end{equation}
where the initial density matrix $\rho_{m}=\prod_{i} e^{-\beta_i[E_{im}-\mu_i N_{im}]}/Z_0^-$. It follows from the antiunitarity of  $\Theta$ that $Z_0^+=Z_0^-$, that is, the Gibbs distribution $\rho_m$
in the time-reversed basis is the same as Eq.~(\ref{gibbs}). One then finds the distribution of the energy and particle number changes, similar to Eq. (\ref{forwarddistri}),
\begin{align}
P[\Delta\mathbf E, \Delta \mathbf N;-] = &\sum_{mn}\prod_{i=1}^r\delta(\Delta E_i - \Delta E_{i,nm})\delta(\Delta N_i - \Delta N_{i,nm}) \nonumber\\ &\times |\langle\psi_n|\Theta^{-1}U(\tau;-)\Theta|\psi_m\rangle|^2\rho_{m}.\label{backwarddistri}
\end{align}
where $\Delta E_{i,nm}=E_{in}-E_{im}$ and $\Delta N_{i,nm}=N_{in}-N_{im}$.

We want to relate the distribution functions (\ref{forwarddistri}) and (\ref{backwarddistri}), i.e., obtain the fluctuation theorem. The evolution operators have an important property\cite{andrieux09}
\begin{equation}
\Theta^{-1}U(\tau;-)\Theta = U^\dag(\tau; +).\label{eq-U}
\end{equation}
This can be seen by checking that both  operators $\Theta^{-1}U(t;-)\Theta$ and $U(\tau-t;+)$ satisfy the equation
\begin{align}
i\frac{d}{dt}V(t) = -\mathcal H(\tau -t;+)V(t). \label{prop}
\end{align}
In terms of  $\Theta^{-1}U(t;-)\Theta$, we find that given the initial condition $V(0)=\Theta^{-1}U(0;-)\Theta=1$, the final operator at $t=\tau$ is $V(\tau) = \Theta^{-1}U(\tau;-)\Theta$. In terms of $U(\tau-t;+)$, given the initial condition $V(0) = U(\tau;+)$, we have $V(\tau) = U(0; +)$. In other words, if we have the initial condition $V(0)=1=U^\dagger(\tau;+)U(\tau;+)$, the final operator at $t=\tau$ will be $V(\tau ) = U^\dag(\tau;+)U(0;+)=U^\dag(\tau;+)$. Therefore, due to the uniqueness of the solution of Eq.~(\ref{prop}), the property (\ref{eq-U}) is obtained.

Now, we combine Eqs.~(\ref{forwarddistri}), (\ref{backwarddistri}) and (\ref{eq-U}), and obtain the fluctuation theorem
\begin{equation}
\frac{P[\Delta\mathbf E, \Delta \mathbf N;+]}{ P[-\Delta\mathbf E, -\Delta \mathbf N;-]}=\prod_{i} e^{\beta_i(\Delta E_i-\mu_i\Delta N_i)} \label{eq-fr}.
\end{equation}
Clearly, it is of a general form (\ref{ftgeneral}). As mentioned above, in the case of time-reversal invariant systems, i.e.,  for $\mathcal H(t;+)=\Theta \mathcal H(\tau -t;+)\Theta^{-1}$, the microscopic dynamical equations (\ref{forwardpropagator}) and (\ref{backwardpropagator}) are the same. Hence, the ``$+$'' and ``$-$'' can be dropped in the fluctuation theorem. However, in what follows, we will focus on systems without the time-reversal symmetry. Therefore, one has to keep in mind that the two distribution functions in (\ref{eq-fr}) describe two different systems. In most of the following applications, the system in the backward process is realized by reversing the direction of the magnetic field $B$ which is present in the system in the forward process.

\section{Saito-Utsumi relations}

The Saito-Utsumi relations \cite{saito} connect transport properties of a conductor in two opposite magnetic fields. They generalize quantum fluctuation relations for the electric current and noise in  time-reversal invariant systems \cite{fl2}.

\begin{figure}[b]
\centering
\includegraphics[width=3in]{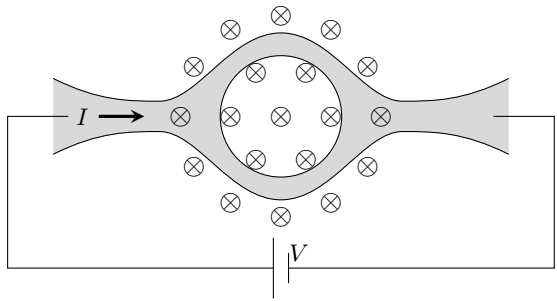}
\caption{ A conductor is placed in a magnetic field and connects two terminals at the same temperature and the voltage difference $V$. }
\label{AAA}
\end{figure}

We consider a conductor, connected to two reservoirs with the voltage difference $V$  and the same temperature $T$, in the presence of a magnetic field, Fig. \ref{AAA}. It is possible to generalize to a multi-terminal case. Below we will only investigate the simplest two-terminal  situation. We will derive the relations between various current correlation functions for two opposite orientations of the magnetic field.  An infinite number of relations can be derived for an infinite number of the correlation functions but we will focus on a few simplest relations that are most relevant for the experiment.

We will be interested in three quantities:

1) The electric current

\begin{equation}
\label{dima1}
I=e\langle\Delta N\rangle/\tau;
\end{equation}

2) The noise power

\begin{equation}
\label{dima2}
S=2e^2(\langle \Delta N^2\rangle-\langle\Delta N\rangle^2)/\tau;
\end{equation}

3) The third cumulant

\begin{equation}
\label{dima3}
C=\frac{2e^3}{\tau}[\langle \Delta N^3\rangle-3\langle\Delta N\rangle\langle\Delta N^2\rangle+2\langle\Delta N\rangle^3],
\end{equation}
where $e<0$ is one electron charge, $\tau$ the duration of the protocol, Section 2, $\Delta N$ the number of electrons transferred between the reservoirs, and the angular brackets denote the average over the distribution $P(\Delta N, B)$, where $B$ is the magnetic field.
Note that $\langle\Delta N\rangle\rightarrow 0$ at $V\rightarrow 0$ since there is no current at zero voltage. We will only compute the noise $S$ with the accuracy up to the linear term in $V$ below. Thus, it is legitimate to omit the contribution $\langle\Delta N\rangle^2$ in Eq. (\ref{dima2}).

The fluctuation relations, considered in this section, were first derived in Ref. \refcite{saito}. We follow a simpler derivation from Ref. \refcite{exp2}.

\subsection{Symmetric and antisymmetric variables}

The results express in the simplest way in terms of the symmetrized and antisymmetrized combinations of the currents and noises $I_{\pm}=I(B)\pm I(-B)$, $S_{\pm}=S(B)\pm S(-B)$ in the opposite magnetic fields $\pm B$.
We introduce the Taylor expansions in powers of the voltage $V$

\begin{equation}
\label{dima4}
I_+=G_1V+\frac{G_2V^2}{2}+\dots;
\end{equation}

\begin{equation}
\label{dima5}
I_-=G_1^AV+\frac{G_2^AV^2}{2}+\dots;
\end{equation}

\begin{equation}
\label{dima6}
S_+=S_0+S_1V+\dots;
\end{equation}

\begin{equation}
\label{dima7}
S_-=S_0^A+S_1^AV+\dots;
\end{equation}

\begin{equation}
\label{dima8}
C_-=C_0^A+\dots.
\end{equation}

The Saito-Utsumi relations connect the above Taylor coefficients.

According to the Onsager reciprocity relations \cite{onsager,casimir}, the linear conductance is an even function of the magnetic field, $G_1(B)=G_1(-B)$. Hence, $G_1^A=0$. Next, the Nyquist formula \cite{nyquist} implies that
the equilibrium noise power $S_1=4G_1T$ does not depend on the direction of the magnetic field and $S_0^A=0$. We will also see that the symmetrized third cumulant $C_+=C(B)+C(-B)$ is zero at $v=0$.

\subsection{Fluctuation relations for symmetric variables}

We will use the notation $P(\Delta N,B)$ for the probability to transfer $\Delta N$ electrons from the left reservoir to the right one during the protocol, Section 2. According to the fluctuation theorem (\ref{eq-fr}),

\begin{equation}
\label{dima9}
P(\Delta N,B)=P(-\Delta N,-B)e^{v\Delta N},
\end{equation}
where $v=eV/T$.
We next introduce the symmetrized and antisymmetrized probabilities $P_{\pm}(\Delta N)=P(\Delta N,B)\pm P(\Delta N,-B)$. Eq. (\ref{dima9}) then yields

\begin{equation}
\label{dima10}
P_{\pm}(\Delta N)=\pm P_{\pm}(-\Delta N)e^{v\Delta N}.
\end{equation}
We also introduce the symmetrized and antisymmetrized averages $\langle\Delta N^k\rangle_\pm=\sum P_{\pm}(\Delta N)\Delta N^k$.

Eq. (\ref{dima10}) for $P_+$ assumes exactly the same form as the fluctuation theorem for $P(\Delta N,B=0)$. Thus, all fluctuation relations for the symmetric currents $I_+$ and noises $S_+$ are the same as the relations for the currents and noises in the absence of the magnetic field.

Note that at $v=0$, Eq. (\ref{dima10}) yields $P_+(\Delta N)=P_+(-\Delta N)$. Hence, $\langle \Delta N^{2k+1}\rangle_+=\sum P_{+}(\Delta N)\Delta N^{2k+1}$ is zero at $v=0$ for any odd $2k+1$.
A trivial consequence of this relation is the absence of the average electric current in equilibrium, $I_+=e\langle \Delta N\rangle/\tau=0$. We also find that $C_+=C(B)+C(-B)=0$.

\begin{figure}[b]
\centering
\includegraphics[width=3in]{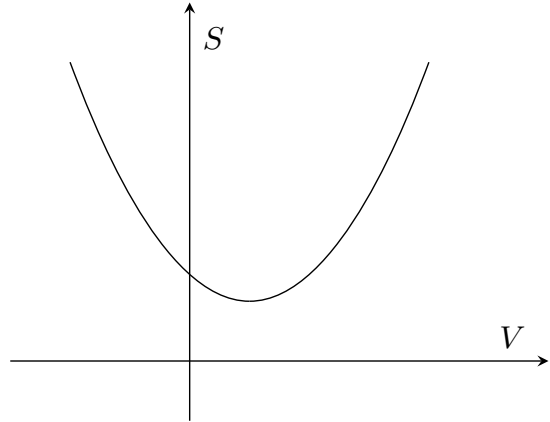}
\caption{Noise is minimal at a nonzero $V$.}
\label{BBB}
\end{figure}

We now expand in powers of $v$ the left and right hand sides of the relation

\begin{equation}
\label{dima11}
\langle\Delta N\rangle_+=\sum_{\Delta N}\Delta N P_+(\Delta N)=-\sum_{\Delta N} \Delta N P_+(\Delta N)e^{-v\Delta N}.
\end{equation}
After defining

\begin{equation}
\label{dima12}
\langle\Delta N^k\rangle_\pm=N^\pm_{k,0}+vN^\pm_{k,1}+\frac{v^2}{2}N^\pm_{k,2}+\dots
\end{equation}
we obtain $N^+_{2,0}=2N^+_{1,1}$ and $N^+_{1,2}=N^+_{2,1}$, where we used the fact that $N^+_{3,0}=0$. Comparing with Eqs. (\ref{dima4}) and (\ref{dima6}), we finally obtain

\begin{equation}
\label{dima13}
S_0=4G_1T;
\end{equation}
\begin{equation}
\label{dima14}
S_1=2TG_2.
\end{equation}
The first equation is nothing but the Nyquist formula. Eq. (\ref{dima14}) goes beyond the standard fluctuation-dissipation relation since it contains the nonlinear transport coefficient $G_2$. Since that coefficient is nonzero in general, the noise $S=S_0+VS_1+\dots$ is minimal at a nonzero voltage, Fig. \ref{BBB}.

\subsection{Fluctuation relations for antisymmetric variables}

We now turn to the new results that distinguish systems in a magnetic field from the time-reversal invariant situation.
First, consider the identity

\begin{equation}
\label{dima15}
\langle \Delta N\rangle_-=\sum_{\Delta N}P_-(\Delta N)\Delta N=\sum_{\Delta N}P_-(\Delta N)\Delta N e^{-v\Delta N}.
\end{equation}
The Taylor expansion of the left and right hand sides yields

\begin{equation}
\label{dima16}
N^-_{3,0}=2N^-_{2,1}.
\end{equation}
This is equivalent to
\begin{equation}
\label{dima17}
S_1^A=\frac{C_0^A}{2T}.
\end{equation}

The normalization of probability implies

\begin{equation}
\label{dima18}
\sum P_-(\Delta N)=0=-\sum P_-(\Delta N)e^{-v\Delta N}.
\end{equation}
One obtains from the expansion of the right hand side in powers of $v$

\begin{equation}
\label{dima19}
3N^-_{1,2}-3N^-_{2,1}+N^-_{3,0}=0.
\end{equation}
This reduces to

\begin{equation}
\label{dima20}
S_1^A-2TG_2^A=C_0^A/3T.
\end{equation}

Equations (\ref{dima17}) and (\ref{dima20}), first derived in Ref. \refcite{saito}, are the main results of Section 3.

\subsection{Microreversibility}

The crucial assumption behind our derivation is microreversibility: we assume that after the magnetic field, the velocities of all particles and their spins are reversed, the system traces its evolution backwards.
As was pointed out in Refs. \refcite{forster08,forster09}, such assumption is counterintuitive in mesoscopic systems without time-reversal symmetry.

\begin{figure}[b]
\centering
\includegraphics[width=3in]{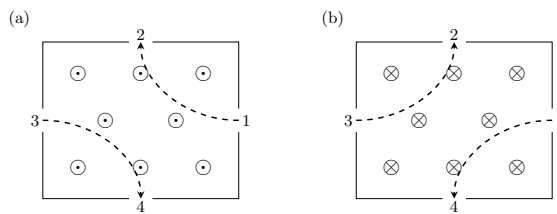}
\caption{Trajectories of charged particles connect different holes in opposite magnetic fields.}
\label{CCC}
\end{figure}

Consider a mesoscopic conductor, connected to several infinite reservoirs, maintained at different voltages. A standard way to calculate currents and noises is based on the Landauer-B\"uttiker formalism \cite{datta}. One first finds the self-consistent charge density $\rho({\bf r})$ as a function of the coordinates in the conductor. The charge distribution creates a self-consistent electrostatic potential. The currents and noises can be computed from the scattering theory for free particles, entering the mesoscopic conductor from the reservoirs, and moving in the self-consistent potential in the conductor. While this picture is compatible with microreversibility at zero magnetic field, there is a conflict at a finite $B$. Indeed, $\rho({\bf r},B)$ is not the same as the charge distribution $\rho({\bf r},-B)$ in the opposite field. Fig. \ref{CCC} illustrates why. In that oversimplified example, charged particles can enter a square box through holes 1 and 3 only. At one direction of the magnetic field, particles from hole 1 move into hole 2 and particles from hole 3 move into hole 4. At the opposite field the particle trajectories connect hole 1 with 4 and 2 with 3. Since the charge density is nonzero only on those trajectories, $\rho({\bf r},B)\ne\rho({\bf r},-B)$. The self-consistent electrostatic potentials depend on $\rho$ and are different in the opposite magnetic fields. Consider now a scattering process in which a particle with the momentum ${\bf k}_i$ acquires the momentum ${\bf k}_f$ at the magnetic field B. The scattering amplitude depends on the self-consistent electrostatic potential $\phi({\bf r},B)$. The scattering probability for the time-reversed process at the field $-B$ describes the transition form the state with the momentum $-{\bf k}_f$ into the state with the momentum $-{\bf k_i}$. It depends on a different self-consistent potential $\phi({\bf r},-B)$ and hence is different from the scattering probability ${\bf k}_i\rightarrow{\bf k}_f$ in the field $B$.
Such asymmetry of the self-consistent potential is closely related to the physics of rectification in mesoscopic conductors \cite{rec1,rec2,rec3,rec4,rec5,rec6,rec7}.

The above argument does not by itself disprove microreversibility. Indeed, it is a mean-field argument which deals with single-particle states. On the other hand, the proof of the fluctuation theorem, Section 2, considers many-particle states. The unitarity of the evolution operator in quantum mechanics shows that the transition probability between the many-body states of the whole system, $P({\rm initial~state~}\rightarrow{\rm~final~state},B)$, always equals $P({\rm time-reversed~final~state~}\rightarrow{\rm~time-reversed~initial~state},-B)$. In each point, the charge densities in the forward and backward processes remain exactly the same in the corresponding moments of time. Hence, the electrostatic potentials do, in fact, remain the same. The mean-field argument is based on the different charge distributions in the steady states in the opposite magnetic fields. However, if one reverses time in a system in a steady state then the charge distribution does not change and hence does not become the steady state  distribution of the electric  charge in the opposite magnetic field.

We would like to emphasize that time-reversal in an interacting macroscopic system is not mathematical fiction. It was demonstrated experimentally long ago in the context of NMR \cite{time1,time2}.

Several groups have verified the fluctuation relations at a finite magnetic field without the use of microreversibility.
The Saito-Utsumi relations were confirmed by microscopic calculations beyond the Landauer-B\"uttiker formalism in two models \cite{saito09,nasb10}. Note also that an approximate calculation beyond the mean-field theory in
Ref. \refcite{lim10} agrees with Eq. (\ref{dima14}). The fluctuation relations for a general chiral system from Section 5 below were derived both from microreversibility \cite{wang2} and without its use \cite{wang1}. Recent experiments \cite{exp1,exp2} were interpreted as supporting the fluctuation relations \cite{saito} (see the next subsection).
While all this gives credibility to the approach, based on microreversibility, one must remember some subtleties.

All models in Refs. \refcite{saito09,nasb10,lim10,wang1} assume a finite range of the electrostatic interaction. This implies the presence of screening gates. The gates are crucial for the fluctuation relations in chiral systems \cite{wang1,wang2} since it is meaningful to speak about chiral transport only in systems with short-range interactions (Section 4). At the same time, the Saito-Utsumi relations do not make assumptions about the range of interactions. Yet, our derivation of the fluctuation theorem, Section 2, implicitly assumes short-range forces. Indeed, in the presence of the long-range Coulomb interaction, the reservoirs are not independent even in the beginning of the forward and backward processes and cannot be described by the Gibbs distribution. Certainly, this is a rather standard issue. It can be resolved by splitting the Coulomb force into a finite-range part with some large but finite interaction radius and the long-range part which must be treated in the mean-field approximation. This allows using the Gibbs distribution but returns the problem from the Landauer-B\"uttiker approach. Indeed, the average charge densities are not the same at the opposite field orientations and hence the mean-field effective long-range potentials are not the same in the forward and backward process. Fortunately, if the reservoirs are sufficiently large and the radius of the finite-range interaction is selected large enough, one can see that the difference of the long-range potentials can be neglected.

Another issue equally affects systems with and without the time-reversal symmetry. Our derivation, Section 2, assumes that the system is isolated. This may not be easy to accomplish in practice. Moreover, if this has been accomplished then no experiments can be performed since any measurement device disturbs the system. Fortunately, energy exchange with the outside world turns out not to be a problem provided the temperature of the environment is the same as the temperature of the system of interest. Indeed, let us include the environment as an additional reservoir and repeat the derivation of Eq. (\ref{eq-fr}). The energy changes $\Delta E_i$  of the reservoirs in the forward process enter Eq. (\ref{eq-fr}) in the form $\sum\beta_i\Delta E_i$. This combination is zero at equal $\beta_i$ from the energy conservation.
Thus, all $\Delta E_i$, including the energy absorbed by the environment, drop out from Eq. (\ref{eq-fr}). On the other hand, the electrostatic potential of the environment drops out only in the absence of the particle exchange with the outside world. In fact, the fluctuation theorem Eq. (\ref{eq-fr}) may break down even if charges are transfered between  different regions of the environment in the absence of the particle exchange with the system of interest
\cite{exp3,exp4}. This issue has been a major difficulty in the experiments on fluctuation relations in mesoscopic conductors as we discuss in the next subsection.

\subsection{Experiment}

Note that Eq. (\ref{dima17}) crucially depends on the microreversibility while Eq. (\ref{dima20}) does not require such assumption as shown in Ref.   \refcite{forster08}, see also Ref. \refcite{s09}. Thus, the verification of Eq. (\ref{dima17}) is particularly interesting. On the other hand, it is easier to measure noises and currents than the third cumulant \cite{reznikov}. As a result, recent experiments have focused on the verification of a consequence of Eqs. (\ref{dima17},\ref{dima20}):

\begin{equation}
\label{dima21}
S_1^A=6TG_2^A.
\end{equation}

Refs. \refcite{exp1,exp2} tested Eqs. (\ref{dima21}) and (\ref{dima14}) in a mesoscopic interferometer. Ref. \refcite{exp1} observed $S_1^A/[TG_2^A]=8.7^{+1.3}_{-0.7}$ and Ref. \refcite{exp2} obtained $S_1^A/[TG_2^A]=9.7^{+1.3}_{-1.2}$ in satisfactory agreement with Eq. (\ref{dima21}). This was interpreted as a proof of microreversibility.
On the other hand, the results for $S_1$, $S_1/TG_2=10.8^{+2.4}_{-1.4}$, Ref. \refcite{exp1}, and $S_1/TG_2=12.0^{+1.9}_{-2.0}$, Ref. \refcite{exp2}, are incompatible with (\ref{dima14}). The reasons for the discrepancy of the theory and experiment are unclear. A related experiment without a magnetic field may provide some hints.

A violation of the fluctuation theorem (\ref{eq-fr}) was found in a single electron tunneling experiment through a double quantum-dot system \cite{exp3,exp4}. The violation has been explained by a careful analysis of the experimental circuit. The circuit included a quantum point contact
(QPC) electrometer, used as a measurement device. When a finite voltage bias is applied to the quantum point contact, the fluctuation theorem (\ref{eq-fr}) must be modified. The right hand side now contains an additional factor $\exp(QV_{QPC}/T)$, where $V_{QPC}$ is  the bias at the QPC and $Q$ is the charge that travels through the QPC during the forward process. The derivation of the modified relation is exactly the same as our argument in Section 2. One just needs to include the QPC, connected to two reservoirs at different electrochemical potentials, into the system under consideration. This interpretation is supported by the fact that a modified fluctuation relation was found to hold in the experiment \cite{exp1,exp2,sinitsyn}.

\section{Chiral systems}

The Saito-Utsumi relations from the previous section are very general and apply to any conductor in a magnetic field. This generality comes at a price. Indeed, the relations connect the coefficients in the expansions of the currents and noises in powers of the voltage. Thus, they only apply at low voltages, that is, close to equilibrium.  One can overcome this limitation
by looking at so-called chiral systems. In such systems all excitations propagate in one direction only, for example all excitations propagate to the right only.

\begin{figure}[t]
\centering
\includegraphics[width=3in]{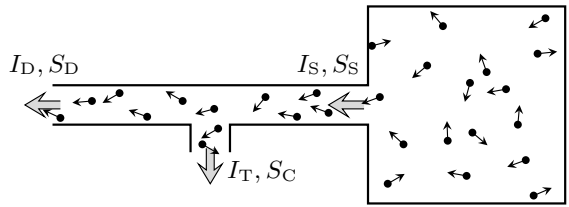}
\caption{Ideal gas in a reservoir with a tube (from Ref. 30).}
\label{idealgas}
\end{figure}

The simplest example of a chiral system is illustrated in Fig. \ref{idealgas}. A box is filled with an ideal gas and placed in vacuum. A narrow tube with an open end and smooth walls is attached to the box. Particles can leave the box through the tube but do not come back. Thus, the transport in the tube is chiral.

The above example is too simple to be interesting other than as a toy model. Several more interesting examples of chiral transport are known. For example, statistical mechanics has been used to model traffic \cite{helbig}. Transport is chiral on a network of one-way roads. In some biological systems, transport can only occur in one direction. Most importantly, chiral transport takes place on the edges and surfaces of some topological states of matter.

\subsection{Chiral transport in topological systems}

Chiral transport occurs in several topological systems without the time-reversal symmetry. For example, the surface of a 3D stack of integer quantum Hall liquids is chiral \cite{3D1,3D2,3D3}. Transport is chiral on the edges of
$p\pm ip$ superconductors \cite{anyons,green,ivanov}.
Besides, many states of the two-dimensional electron gas in the conditions of the quantum Hall effect \cite{perspectives} (QHE) are chiral. So far, most research on chiral transport has been focused on those QHE states.

We expect that the bulk of a quantum Hall system is gapped (see, however, Refs. \refcite{4-3,heiblum14}). Thus, low-energy transport occurs on the edges. Wen's hydrodynamic theory \cite{wen} predicts that in many cases the low-energy transport is chiral.
The integer QHE with the filling factor $\nu=1$ is the simplest example. Wen's theory describes the edge physics in terms of a single field $\phi$, where $\partial_x\phi$ is proportional to the linear charge density. The action assumes the form

\begin{equation}
\label{dima22}
L=\int dxdt[\partial_t\phi\partial_x\phi-v(\partial_x\phi)^2].
\end{equation}
The solution of the equation of motion, $\phi=\phi(x+vt)$,  describes excitations that move only to the left.

Generalizations of the action (\ref{dima22}) also predict \cite{chang,wen} chiral transport at all other integer filling factors and in many fractional QHE states, including the states with the filling factors $\nu=1/3$ and $\nu=2/5$.
Some other QHE states are not expected to be chiral. This point can be easily understood by considering $\nu=2/3$. The $\nu=2/3$ liquid  can be described as the $\nu=1/3$ state of holes on top of the $\nu=1$ state of electrons.
Thus, its edge theory contains two excitation branches, corresponding to $\nu=1$ and $\nu=1/3$, with the opposite chirality of the electron and hole edges.

Such description of the $2/3$ edge conflicts with the experiment that shows only one propagation direction for charged excitations. This was explained by Kane, Fisher and Polchinsky \cite{kfp} who uncovered the nontrivial role of impurities which are inevitably present along QHE edges. Impurities promote tunneling between the contra-propagating edge channels and change the nature of the edge modes: all charge excitations move in the same direction, called downstream; in addition, a neutral excitation branch of the opposite, i.e., upstream, chirality emerges.

A similar picture is expected to apply in several other QHE states. Detecting upstream neutral modes proved to be a great challenge and only recently has progress been reported in the field \cite{36,deviatov,yacoby}. We will see that the fluctuation relations from Section 5 give a tool to test the presence of upstream neutral modes. Confusingly, there were recent reports of upstream modes in the $1/3$ and $2/5$ states and even at the integer filling factors \cite{heiblum14,deviatov,yacoby}. Thus, an independent test of chirality on the edges is of great importance.

\subsection{Non-Abelian quantum Hall states}

The question of chirality on QHE edges also touches upon the ongoing search for non-Abelian anyons \cite{anyons}. Quasiparticles in fractional QHE states are known to be anyons with a different exchange statistics from bosons and fermions \cite{wen}. The simplest Abelian anyons accumulate nontrivial phases when encircle each other. The many-body quantum state of a system of Abelian anyons does not change after one of them makes a circle around another adiabatically. Hypothetical  non-Abelian anyons \cite{anyons,2}  change their quantum state after one particle braids another. This does not involve a change of the internal quantum numbers of any particle and reflects the fact that the information about a quantum state of an anyonic system is distributed over the whole system. This property makes non-Abelian anyons attractive for topological quantum information processing, naturally protected from errors \cite{anyons,6}. Indeed, local perturbations from the interaction with the environment cannot change or erase quantum information that is stored globally.

The most promising candidate for non-Abelian anyons is the QHE state \cite{willett87} at $\nu=5/2$. At the same time, there are many competing Abelian and non-Abelian candidate states at that filling factor  \cite{2,3,4,5,8,19,BS,26,Overbosch,yang} and the existing body of experiments does not allow the determination of the right state at this point, see Ref. \refcite{yang} for a review. Interferometry
 \cite{Stern10,chamon97,fradkin98,mz,dassarma05,11,12,13,14,hou06,grosfeld06,15,willett09,bishara09,willett10,fp331,16,kang,double,rosenow12} is the most direct approach but its implementation has faced significant difficulties. This motivated the search for non-interferometric ways to obtain information about the $5/2$ state
 \cite{yang,feldman08,viola12,overbosch09,CS,seidel09,yang09,wang10a,mstern10,rhone11,tiemann12,mstern12,chickering13,20,21}.
Neither non-interferometric method would provide a direct observation of anyonic statistics but their combination may be sufficient to identify the correct state. In particular, some of the proposed states are chiral while others are not. This makes a chirality test, based on the fluctuation relations from Section 5, a useful tool in the search for non-Abelian particles.

\subsection{Chirality and causality}

In chiral systems transport occurs in one direction only. This includes the transport of information. Thus, the causality principle is enhanced. In addition to the requirement that future events do not affect the past, we also expect that the downstream events do not affect upstream events even in the future. Such modified causality principle allows one to generalize the Nyquist formula for nonlinear transport far from equilibrium (Section 5). Indeed, the standard derivation of the fluctuation-dissipation theorem (FDT)  is based on the combination of the Gibbs distribution and linear response theory. Causality is the key ingredient of the latter theory. The enhanced causality principle allows a derivation of a generalized FDT even without the use of the Gibbs distribution, that is, far from equilibrium.

Note that our definition of chirality assumes the absence of long-range forces \cite{wang1}. Otherwise, such forces could mediate instantaneous information exchange between distant points so that the upstream events would affect the events downstream. In particular, we assume the presence of screening gates in chiral systems of charged particles.

\section{Fluctuation relations in chiral systems}

In this section we focus on chiral systems. We discover that fluctuation relations assume a form \cite{wang1,wang2}, similar to the equilibrium FDT, but hold for nonlinear transport away from equilibrium.
We start with  a heuristic derivation in Subsection 5.1. In Subsection 5.2 we verify the nonequilibrium FDT in the toy model, Fig. ~\ref{idealgas}. We then give a general proof in Subsection 5.3. Numerous generalizations \cite{wang2} and possible applications are briefly addressed in Subsections 5.4 and 5.5.

One closely related fluctuation relation was derived in Refs. \refcite{kf,FLS,FS} in an exactly solvable model. Interestingly, that model could be used to describe both a chiral system in the context of QHE physics and a nonchiral quantum wire.
Our results show that the integrability of the Hamiltonian is not required for the existence of an infinite number of fluctuation relations in chiral systems. At the same time, chirality is crucial. The applicability of the results of Refs. \refcite{kf,FLS,FS} in a nonchiral system is a peculiar feature of the exactly solvable model.

\subsection{Qualitative argument}

\begin{figure}[b]
\centering
\includegraphics[width=3in]{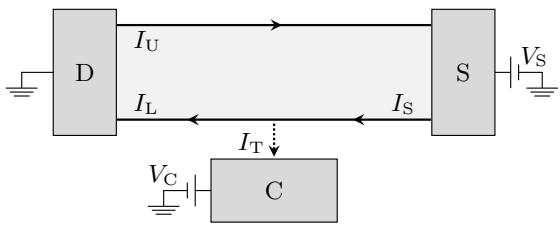}
\caption{ From Ref. 30. Three-terminal setup of a chiral system. We consider a quantum Hall bar with the lower edge, coupled to terminal C. Charge can tunnel  between terminal C and the lower edge. The
 solid lines represent chiral edge modes whose directions are shown by arrows and determined by the external magnetic field. The arrow on the dotted line represents our
convention about the positive direction of the tunneling current $I_{\rm T}$. }
\label{3terminal}
\end{figure}

In the next three subsections, we only consider the simplest example of the nonequilibrium FDT\cite{wang1,wang2}. The extended causality principle (Subsection 4.3)
will be extensively used in the derivation of these results. Many more generalized nonequilibrium fluctuation relations, e.g, fluctuation relations in a
multi-terminal setup and for heat transport, can be derived\cite{wang2}  but their discussion is postponed to section 5.4.

Let us study the nonequilibrium FDT in the three-terminal system shown in Fig.~\ref{3terminal}. It is a Hall bar with one of the two chiral edges coupled to a
third terminal. The three terminals [the source (S), the drain (D) and the third contact (C)] are assumed to be ideal reservoirs that have zero impedance
and infinitely large capacitance. The bulk of the Hall bar is gapped and the edges are chiral. The strength of the coupling
between the Hall bar and the third contact is unimportant. We assume that the two chiral edges are fully absorbed by  source S at the left end
and  drain D at the right end, and they are far apart so that they do not interfere with each other.
Source S is biased at the  voltage $V_{\rm S}$,  contact C is at $V_{\rm C}$, and the drain is grounded. Let the
temperature of the system be $T$. Steady currents flow between the terminals, in particular, charge tunnels into terminal C, leading to a tunneling current
$I_{\rm T}$. We will show that the following relation holds
\begin{equation}
\label{chenjie1}
S_{\rm D} = S_{\rm C} - 4T\frac{\partial I_{\rm T}}{\partial V_{\rm S}} + 4GT,
\end{equation}
where $S_{\rm D}=\int dt \langle \Delta I_{\rm D}(t) \Delta I_{\rm D}(0) + \Delta I_{\rm D}(0) \Delta I_{\rm D}(t)\rangle$ is the zero-frequency noise of
the drain current $I_{\rm D}$, $S_{\rm C}=\int dt \langle \Delta I_{\rm T}(t) \Delta I_{\rm T}(0) + \Delta I_{\rm T}(0) \Delta I_{\rm T}(t)\rangle$ is the
zero-frequency noise of the tunneling current $I_{\rm T}$, and $G$ is the quantized Hall  conductance in the absence of  contact C. The FDT (\ref{chenjie1})
holds for arbitrary $T$, $V_{\rm S}$ and $V_{\rm C}$ as long as they are far below the QHE gap. Note that the system is far from equilibrium when $T
\lesssim V_{\rm C}, V_{\rm S}$.

Below we give a heuristic derivation of the nonequilibrium FDT (\ref{chenjie1}) following Ref.~\refcite{wang1}. As shown in Fig.~\ref{3terminal},
the current $I_{\rm D} = I_{\rm L} - I_{\rm U}$, where $I_{\rm L}$ is the current, entering the drain along the lower edge, and $I_{\rm U}$ is the current,
emitted from the drain along the upper edge. Since the system is in a steady state and no charge is accumulated on the lower edge, the low-frequency part
of $I_{\rm L}$ can be written as $I_{\rm L} = I_{\rm S} - I_{\rm T}$, where $I_{\rm S}$ is the current, emitted from the source. Because the two edges are
uncorrelated, the low-frequency noises obey the relation
\begin{equation}
\label{chenjie2}
S_{\rm D} = S_{\rm C} -2 S_{\rm ST} + S_{\rm S} + S_{\rm U},
\end{equation}
where $S_{\rm ST}=\int dt \langle \Delta I_{\rm T}(t) \Delta I_{\rm S}(0) + \Delta I_{\rm S}(0) \Delta I_{\rm T}(t)\rangle$ is the cross noise of $I_{\rm
S}$ and $I_{\rm T}$, and $S_{\rm S}$ and $S_{\rm U}$ are the noises of $I_{\rm S}$ and $I_{\rm U}$ respectively.  To derive (\ref{chenjie1}) from
(\ref{chenjie2}), let us first find $S_{\rm S}$ and $S_{\rm U}$. To do this, it is enough to consider a simplified case, where contact C is absent.
In that case, we have an obvious result: both $S_{\rm S}$ and $S_{\rm U}$ are equal to one half of the standard Nyquist noise, that is
\begin{equation}
\label{chenjie3}
S_{\rm S}= S_{\rm U}= 2GT.
\end{equation}
Crucially, in the presence of contact C, Eq.(\ref{chenjie3}) still holds. We notice that adding contact C does not affect the noise of $I_{\rm S}$
because of the extended causality principle in chiral systems. Also, it does not affect the noise of $I_{\rm U}$ because of the assumption that the two edges are
uncorrelated. Hence, the noises $S_{\rm S}$ and $S_{\rm U}$ are given by Eq. (\ref{chenjie3}) even in the presence of terminal C.

We are left with the calculation of the cross noise $S_{\rm ST}$. Let us analyze the dependence of the tunneling current $I_{\rm T}(t)$ on the emitted
current $I_{\rm S}$. The tunneling current $I_{\rm T}$ depends on the average emitted current $\bar I_{\rm S}$ and its fluctuations $I_{\omega}$, where
$\omega$ denotes the fluctuation frequency. According to the extended causality principle, the average emitted current $\bar I_{\rm S}$
depends only on $V_{\rm S}$ and is not affected by terminal C. In other words,  $\bar I_{\rm S}= G V_{\rm S}$.  We now assume that the central part of the lower edge has a relaxation time $\tau$, so that
the instantaneous value of $I_{\rm T}(t)$ depends only on the emitted current within the time interval $\tau$. It is convenient to separate the
fluctuations of $I_{\rm S}$ into the fast part $I^>$ which contains the fluctuations of the  frequencies above $1/\tau$, and the slow part $I^<$ which contains
the fluctuations of the frequencies below $1/\tau$. Within the time interval $\tau$, $I^<$ does not exhibit a time-dependence. Hence $I^<$ enters the expression for
the tunneling current $I_{\rm T}(t)$ in the combination $\bar I_{\rm S}+I^<$ only, $I_{\rm T} = \langle I(\bar I_{\rm S}+I^<, I^>)\rangle$, where the
brackets denote  the  average with respect to the fluctuations of $I_{\rm S}$. According to the Nyquist formula for the emitted current, its harmonics with
different frequencies have zero correlations $\langle I_{\omega} I_{-\omega'} \rangle\sim \delta(\omega-\omega')$. For the sake of the heuristic argument,
we will assume a Gaussian distribution of $I_{\rm S}$, hence, independence of the high- and low-frequency fluctuations. With this assumption, we
average over the fast fluctuations and write $I_{\rm T} = \langle J(\bar I_{\rm S}+I^<)\rangle$, where $J$ is obtained by averaging over $I^>$.  $I^<$
corresponds to a narrow frequency window and can be neglected in comparison with $\bar I_{\rm S}$, i.e., the average tunneling current $I_T= J(\bar I_{\rm
S})$. For the calculation of the cross noise, we expand $J(\bar I_{\rm S}+I^<)$ to the first order in $I^<$ and obtain
\begin{equation}
\label{chenjie4}
S_{\rm ST} = \langle  I_{\rm T,\omega}I_{-\omega} + I_{\rm T, -\omega} I_{\omega}\rangle = \frac{\delta J(\bar I_{\rm S})}{\delta \bar I_{\rm S}} \langle
I_\omega I_{-\omega} + I_{-\omega} I_{\omega}\rangle = 2T \frac{\partial I_{\rm T}}{\partial V_{\rm S}},
\end{equation}
where we have used the result $S_{\rm S} =\langle I_\omega I_{-\omega} + I_{-\omega} I_{\omega}\rangle=2GT$, and $\delta \bar I_{\rm S}= G\delta V_{\rm
S}$. Combining the results (\ref{chenjie2})-(\ref{chenjie4}), the nonequilibrium FDT (\ref{chenjie1}) is easily obtained.

We have seen that chirality plays an important role in the derivation of Eqs. (\ref{chenjie3}) and (\ref{chenjie4}). In this heuristic derivation, an unnecessary
assumption of Gaussian fluctuations of $I_{\rm S}$ has been made. In Subsection 5.3, the nonequilibrium FDT will be derived from the fluctuation theorem (\ref{eq-fr}) without that
assumption.

\subsection{Toy model}

Before moving  to a general proof of the nonequilibrium FDT (\ref{chenjie1}), let us verify it in a toy model of an ideal gas
(Fig.~\ref{idealgas}). Our discussion will follow the appendix to Ref. \refcite{wang1}.

Consider a large reservoir of an ideal gas of noninteracting molecules at the temperature $T$ and chemical potential $\mu$.
Molecules can leave the reservoir through a narrow tube with smooth walls. By smooth, we mean such walls that the  collisions of the molecules with the walls are elastic and
do not change the velocity projection along the tube axis. Thus, molecules can only leave the reservoir but never come back. The system is chiral. Imagine
now that molecules can escape through a side  hole in the wall of the tube (Fig. ~\ref{idealgas}). We can derive a relation, similar to Eq.~(\ref{chenjie1}):
\begin{equation}
\label{chenjie6}
S_{\rm D} = S_{\rm C} - 4T \frac{\partial I_{\rm T}}{\partial \mu} + S_{\rm S},
\end{equation}
where $I_{\rm T}$ is the particle current through the side hole in the tube wall and $S_{\rm C}$ is its noise, Fig. ~\ref{idealgas}, $S_{\rm D}$ is the noise of the current $I_{\rm
D}$ at the open end of the tube, and $S_{\rm S}$ is the noise of the particle current $I_{\rm S}$ at the opposite end, attached to the box. The noise $S_{\rm S}$ can be determined
from the measurement of $S_{\rm D}$ in the absence of the side hole in the tube wall. Note that the above relation is almost the same as the nonequilibrium FDT
(\ref{chenjie1}) in the QHE setup, Fig.~\ref{3terminal}.

The proof of the above expression is rather simple and builds on Ref.~\refcite{martin}.
It is most convenient to work with a Fermi gas with
a high negative chemical potential.
Other cases, such as a Bose gas, can be considered in a similar way but will not be addressed below.
Let $f=1/\{\exp[(E-\mu)/T]+1\}$ be the Fermi distribution in the box and $T_{E}$ the transmission
coefficient through the side hole in the tube wall for a particle of the energy $E$. According to Ref.~\refcite{martin}, for the particles within the energy window $(E,
E+dE)$, the current through the side hole is $T_{E}f dE$ and the noises $S_{\rm S}=2f(1-f)dE$, $S_{\rm C} = 2T_{E}f(1-T_{E}f)dE$, and $S_{\rm
D}=2(1-T_{E})f[1-(1-T_{E})f]dE$. One  needs to integrate over the energy to obtain the overall current and noises. It is then easy to obtain the
expression (\ref{chenjie6}).

\subsection{General derivation}

\begin{figure}[b]
\centering
\includegraphics[width=3in]{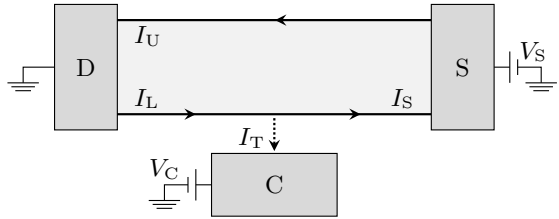}
\caption{ The time-reversed setup. The only differences from Fig.~\ref{3terminal} are the reversed directions of the magnetic field and the chiral edge modes.}
\label{3terminal-tr}
\end{figure}

In this subsection, we give a derivation of the nonequilibrium FDT (\ref{chenjie1}) based on the fluctuation theorem (\ref{eq-fr}).  In the
heuristic argument, Subsection 5.1, an unnecessary assumption was only made in the derivation of the expression for the cross noise (\ref{chenjie4}). Thus, below we
focus on proving the relation (\ref{chenjie4}).

Fig.~\ref{3terminal} is a three-terminal case of the most general setup Fig.~\ref{fig1}, so the fluctuation theorem (\ref{eq-fr}) applies. Let us follow
the protocol in Section 2 and assume that after the time period $\tau$, the changes of the particle numbers in the three terminals are $\Delta N_{\rm S}$,
$\Delta N_{\rm C}$, and $\Delta N_{\rm D} = -\Delta N_{\rm S} - \Delta N_{\rm C}$, respectively, and the distribution function is $P(\Delta N_{\rm S},
\Delta N_{\rm C}; B)$. Note that only two out of the three $\Delta N$'s are independent due to the charge conservation. Then
the fluctuation theorem (\ref{eq-fr}) becomes

\begin{equation}
\label{chenjie7}
\frac{P(\Delta N_{\rm S}, \Delta N_{\rm C};+ B)}{P(-\Delta N_{\rm S}, -\Delta N_{\rm C}; -B)}= e^{-\beta e(V_{\rm S}\Delta N_{\rm S} + V_{\rm C} \Delta
N_{\rm C})}
\end{equation}
where $\beta =1/T$, $e$ is the electron charge,  and $P(\Delta N_{\rm S}, \Delta N_{\rm C};-B)$ is the distribution function for the backward process (see Section 2) in the time-reversed version
of Fig.~\ref{3terminal} as illustrated in Fig. ~\ref{3terminal-tr}. Note that we could write the drain voltage $V_{\rm D}$ explicitly in the fluctuation theorem instead
of setting $V_{\rm D}=0$. That, however, would only burden our notation without providing any advantage. Fig.~\ref{3terminal-tr} shows the
time-reversed setup which has the opposite chirality, compared to Fig. ~\ref{3terminal}. It is worth emphasizing that the observables $I_{\rm T}$, $S_{\rm ST}$, {\it etc.}, that we are
interested in are defined in the setup from Fig.~\ref{3terminal}. The role of the time-reversed setup is only to help us prove the nonequilibrium
FDT (\ref{chenjie1}). In terms of the distribution function $P(\Delta N_{\rm S}, \Delta N_{\rm C}; B)$, the observables of interest can be written as

\begin{equation}
\quad I_{\rm T}   =\frac{e}{\tau}\langle\Delta N_{\rm C}\rangle,
\end{equation}

\begin{equation}
I_{\rm R} = I_{\rm U} - I_{\rm S}= \frac{e}{\tau}\langle\Delta N_{\rm S}\rangle,
\end{equation}

\begin{equation}
S_{\rm ST} = -S_{\rm R T} = -2\frac{e^2}{\tau}(\langle\Delta N_{\rm S}\Delta N_{\rm C}\rangle - \langle\Delta N_{\rm S}\rangle\langle\Delta N_{\rm
C}\rangle),
\end{equation}
where
\begin{equation}
\langle x \rangle = \sum_{\Delta N_{\rm S}, \Delta N_{\rm C}} x P(\Delta N_{\rm S}, \Delta N_{\rm C}; B).
\end{equation}
We have defined the current $I_{\rm R}$ as the overall current flowing into source S. The cross noise $S_{\rm RT}$ of $I_{\rm R}$ and $I_{\rm T}$
equals $-S_{\rm ST}$ since $I_{\rm U}$ is uncorrelated with $I_{\rm T}$.

It is convenient to define the cumulant generating functions

\begin{equation}
\label{chenjie8}
Q(x,y;\pm B) = \lim_{\tau\rightarrow \infty} \frac{1}{\tau} \ln\left\{\sum_{\Delta N_{\rm S}, \Delta N_{\rm C}} e^{-x e\Delta N_{\rm S} - y e\Delta N_{\rm
C}} P(\Delta N_{\rm S}, \Delta N_{\rm C};\pm B)\right\}.
\end{equation}
With these generating functions, the quantities of interest can be expressed as

\begin{equation}
\label{def1}
I_{\rm R} = -\left.\frac{\partial Q(x,y; +B)}{\partial x}\right|_{x= y= 0},
\end{equation}

\begin{equation}
\label{def2}
I_{\rm T} = -\left.\frac{\partial Q(x,y;+B)}{\partial y}\right|_{x=y=0},
\end{equation}

\begin{equation}
\label{def3}
S_{\rm ST} = -2\left.\frac{\partial^2 Q(x,y; +B)}{\partial x \partial y}\right|_{x=y=0}.
\end{equation}

Inserting the fluctuation theorem (\ref{chenjie7}) into the definition (\ref{chenjie8}), we find that the generating functions $Q(x,y;\pm B)$ have a very
nice property:
\begin{equation}
\label{chenjie5}
Q(x, y; +B) = Q(-\beta V_{\rm S} - x, -\beta V_{\rm C} -y; -B).
\end{equation}
Note that this is an equation, relating two generating functions, $Q(x, y; +B)$ and $Q(x, y; -B)$. Also, because the distribution functions $P(\Delta
N_{\rm S}, \Delta N_{\rm C};\pm B)$ are normalized, we have $Q(x=0, y=0; \pm B)=0$.  Since the distribution functions  depend on the biases $V_{\rm S}$
and $V_{\rm C}$, the generating functions are also functions of $V_{\rm S}$ and $V_{\rm C}$
\begin{equation}
Q = Q(x, y, V_{\rm S}, V_{\rm C}; \pm B).
\end{equation}

 We now prove the relation (\ref{chenjie4}) and hence also the nonequilibrium FDT
(\ref{chenjie1}) from the property (\ref{chenjie5}) and enhanced causality.  We first apply the differential operator

\begin{equation}
\hat D = \left(\frac{d}{dx} - T \frac{d}{d V_{\rm S}}\right)\left(\frac{d}{dy} - T \frac{d}{d V_{\rm C}}\right)
\end{equation}
on both sides of Eq.~(\ref{chenjie5}). A straightforward calculation using the expressions (\ref{def1})-(\ref{def3}) yields

\begin{align}
\label{chenjie9}
\frac{1}{2}S_{\rm ST} = & T\frac{\partial I_{T}}{\partial V_{\rm S}} +T\frac{\partial I_{R}}{\partial V_{\rm C}}+  T^2\left.\frac{\partial^2Q(x, y, V_{\rm
S}, V_{\rm C};+B)}{\partial V_{\rm S}\partial V_{\rm C}}\right|_{x=y=0} \nonumber \\
& -T^2 \left.\frac{\partial^2Q(x, y, V_{\rm S}, V_{\rm C};-B)}{\partial V_{\rm S}\partial V_{\rm C}}\right|_{x=-\beta V_{\rm S}, y=-\beta V_{\rm C}}.
\end{align}

\noindent
This equation does not look the same as Eq.~(\ref{chenjie4}), which is simply $S_{\rm ST} = 2T\frac{\partial I_{T}}{\partial V_{\rm S}} $. However, we will
show that only the first term on the right-hand-side of Eq.~(\ref{chenjie9}) is nonzero.

The third term on the right-hand-side of Eq.~(\ref{chenjie9}) is
zero simply because $Q(x=0, y=0, V_{\rm S}, V_{\rm C}; B)=0$.

Now the chirality-enhanced causality principle enters the game as a key tool to prove that the second and fourth terms  vanish. First, it is easy to
see that the second term is zero, because: (1) $I_{\rm R} = I_{\rm U} - I_{\rm S}$; (2) $I_{\rm S}$ does not depend on $V_{\rm C}$ due to extended causality and
(3) the upper and lower edges are assumed to be uncorrelated, so $I_{\rm U}$ does not depend on $V_{\rm C}$.

To prove that the last term is zero, we need a more sophisticated argument from  extended causality. Note that the last term comes from the
time-reversed setup, Fig.~\ref{3terminal-tr}. Since the upper and lower edges in Fig.~\ref{3terminal-tr}  are uncorrelated, we can write the
distribution function $P(\Delta N_{\rm S}, \Delta N_{\rm C};- B)$ as

\begin{equation}
P(\Delta N_{\rm S}, \Delta N_{\rm C};- B)= \sum_{\Delta N_{\rm S}'} P_1(\Delta N_{\rm S}';- B) P_2(\Delta N_{\rm S}-\Delta N_{\rm S}', \Delta N_{\rm C};-
B),
\end{equation}
where $P_1(\Delta N_{\rm S}';- B)$ is the probability that $-\Delta N_{\rm S}'$ particles  leave  source $S$ through the {\it
upper} edge during the time interval $\tau$, and $P_2(\Delta N_{\rm S}-\Delta N_{\rm S}', \Delta N_{\rm C};- B)$ is the probability that $\Delta
N_{\rm S}-\Delta N_{\rm S}'$  particles enter S through the {\it lower} edge while $\Delta N_{\rm C}$  particles  enter  contact
C. This property of the distribution function $P(\Delta N_{\rm S}, \Delta N_{\rm C};- B)$ allows one to decompose the generating function $Q(x,y;- B)$ as

\begin{equation}
\label{dimaQ}
Q(x,y;- B) = Q_1(x,y;- B)+Q_2(x,y;- B),
\end{equation}

\begin{equation}
Q_1(x,y;- B) = \lim_{\tau\rightarrow \infty} \frac{1}{\tau} \ln\left\{\sum_{\Delta N_{\rm S}'} e^{-x e\Delta N_{\rm S}'} P_1(\Delta N_{\rm S}';-
B)\right\},\nonumber
\end{equation}

\begin{equation}
Q_2(x,y;- B) = \lim_{\tau\rightarrow \infty} \frac{1}{\tau} \ln\left\{\sum_{\Delta N_{\rm S}'', \Delta N_{\rm C}} e^{-x e\Delta N_{\rm S}'' - y e\Delta
N_{\rm C}} P_2(\Delta N_{\rm S}'', \Delta N_{\rm C};- B)\right\}. \nonumber
\end{equation}
With the above decomposition (\ref{dimaQ}), we look at the dependences of $Q_1$ and $Q_2$ on the voltages $V_{\rm S}$ and $V_{\rm C}$. Since the upper edge is chiral, the
distribution $P_1$ depends on $V_{\rm S}$ but not on $V_{\rm C}$. Similarly, since the lower edge is chiral, the distribution $P_2$ does not depend on
$V_{\rm S}$ while it does depend on $V_{\rm C}$. Therefore, $Q_{1}$ depends on $V_{\rm S}$ but not on $V_{\rm C}$, while $Q_{2}$ depends on $V_{\rm C}$
but not on $V_{\rm S}$. Hence,
\begin{equation}
\frac{\partial^2Q(x, y, V_{\rm S}, V_{\rm C};-B)}{\partial V_{\rm S}\partial V_{\rm C}}= \frac{\partial^2Q_1}{\partial V_{\rm S}\partial V_{\rm C}}
+\frac{\partial^2Q_2}{\partial V_{\rm S}\partial V_{\rm C}} =0.
\end{equation}
Thus, the last term on the right-hand-side of Eq.~(\ref{chenjie9}) is zero.

To sum up, we have proved that three of the terms on the right-hand-side of Eq.~(\ref{chenjie9}) are zero, leaving the equation to be simply
\begin{equation}
S_{\rm ST} = 2 T\frac{\partial I_{T}}{\partial V_{\rm S}},
\end{equation}
which is exactly the relation (\ref{chenjie4}). Therefore, the nonequilibrium FDT (\ref{chenjie1}) is indeed true for chiral systems.

\subsection{Generalizations}

Above, we studied the simplest example of the nonequilibrium FDT for charge transport. Let us mention some generalizations.

One of the generalizations is to study chiral heat transport in the three-terminal setup in Fig.~\ref{3terminal}. In this case, the three terminals are at different temperatures, $T_{\rm S}$ in S, $T_{\rm C}$ in C, and $T_{\rm D}$ in D. We can assume that they have the same chemical potential.  According to Ref.\refcite{wang2}, the cross noise $S_{\rm ST}^h$ of the heat currents $I_{\rm S}^h$ and $I_{\rm T}^h$ satisfies
\begin{equation}
S_{\rm ST}^h =2T_{\rm S}^2 \frac{\partial I^h_{\rm T}}{\partial T_{\rm S}},
\end{equation}
where $I_{\rm T}^h$ is the heat current flowing into contact C and $I^h_{\rm S}$ is the heat current flowing out of source S. This expression, valid for nonequilibrium states, resembles the standard equilibrium FDT which has the form $S^h = 4T^2 G^h$ with $S^h$ being the noise of the heat current in equilibrium, $G^h$ being the thermal conductance and $T$ the temperature.

Nonequilibrium  fluctuation relations of higher-order cumulants can also be obtained in chiral systems. For example, the following relation holds\cite{wang2} for the three-terminal setup in Fig.~\ref{3terminal}
\begin{equation}
\label{chenjie10}
C_{\rm TT S} = T\frac{\partial S_{\rm TT}}{\partial V_{\rm S}},
\end{equation}
where $S_{\rm TT}$ is the noise of the current $I_{\rm T}$, and $C_{\rm TTS}$ is the third cumulant  defined as
\begin{equation}
C_{\rm TTS} = -\frac{2e^3}{\tau}\langle(\Delta N_{\rm C}-\langle\Delta N_{\rm C}\rangle)\cdot(\Delta N_{\rm C}-\langle\Delta N_{\rm C}\rangle) \cdot (\Delta N_{\rm S}-\langle\Delta N_{\rm S}\rangle)\rangle.
\end{equation}

It is also possible to generalize nonequilibrium fluctuation relations to multi-terminal systems and to higher-order cumulants for heat transport. The reader may consult Ref.~\refcite{wang2} for details.

\subsection{Applications}

Our main result, Eq. (\ref{chenjie1}), as well as its generalization from the previous subsection apply to chiral systems only. Thus, the nonequilibrium FDT can be used to test edge chirality. If the FDT is satisfied both in and beyond equilibrium this is compatible with chirality. If it is broken then the edge is not chiral.

\begin{figure}[t]
\centering
\includegraphics[width=3in]{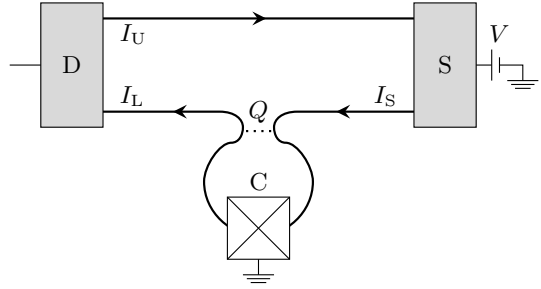}
\caption{From Ref. 30. A possible experimental setup. Charge carriers, emitted
from the source, can either tunnel through constriction Q and
continue toward the drain or are absorbed by  Ohmic contact C.}
\label{EEE}
\end{figure}

\begin{figure}[t]
\centering
\includegraphics[width=3in]{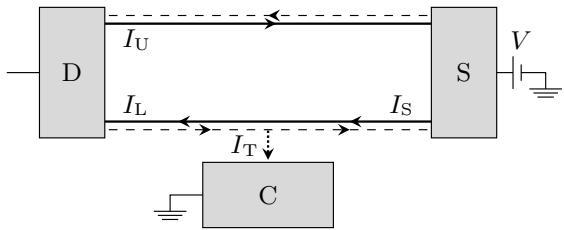}
\caption{From Ref. 30. A nonchiral system. The solid line along the lower edge
illustrates the downstream mode, propagating from the source to
the drain. The dashed line shows a counter-propagating upstream
mode}
\label{DDD}
\end{figure}

A possible experimental setup is shown in Fig. \ref{EEE}.
One of the mechanisms how our FDT gets broken in nonchiral systems is illustrated in Fig. \ref{DDD}. The downstream charged mode dissipates energy in the hot spot \cite{klass},  where it enters the drain. The upstream neutral mode carries the dissipated energy back to the tunneling contact and the point, where the charged mode exits the drain, and heats them. This affects both the tunneling current and the noise and breaks the theorem (\ref{chenjie1}).

Recently there has been much interest in neutral modes on QHE  edges. They were first reported \cite{36} in the particle-hole conjugated QHE states, where the theory predicts upstream neutral modes. Latter, upstream neutral modes were also reported in some states where they had not been expected \cite{heiblum14,deviatov,yacoby}. In such situation a new experimental test, based on the FDT (\ref{chenjie1}), will be helpful.

Our results can also be used to narrow the range of the candidate QHE states at the filling factor $5/2$. Some candidates (e.g., the Pfaffian state \cite{2} or the 331 state \cite{8}) have chiral edges. Others, most notably the anti-Pfaffian state \cite{3,4}, are not chiral.
Some evidence of an upstream neutral mode on the $5/2$ edge has been reported recently \cite{36}. Obviously, verification with a different method is highly desirable. Our FDT (\ref{chenjie1}) provides such a method.

\section{Conclusion}

In this review we considered fluctuation relations in the absence of the time-reversal symmetry. The Saito-Utsumi relations\cite{saito}  apply to any conductor in a magnetic field and connect nonlinear transport coefficients in the opposite magnetic fields close to equilibrium. The fluctuation theorem for chiral systems \cite{wang1,wang2} applies even far from equilibrium and connects currents and noises at the same direction of the magnetic field. This relation can be used to test the chirality of QHE edges. This is relevant in the ongoing search for neutral modes on QHE edges and in the search for non-Abelian anyons.

Many questions remain open. New experimental tests of the fluctuation relations \cite{saito} beyond Refs. \refcite{exp1,exp2} would be important. The conflict between the existing  experimental data and Eq. (\ref{dima14}) has not been understood yet.
On the theory side, fluctuation relations for electric currents can be generalized for any other conserving quantity. In particular, Ref. \refcite{wang2} addresses fluctuation relations for heat currents in chiral systems. Fluctuation relations for spin currents is another interesting question.
In that context time-reversal symmetry may be broken by an external magnetic field or by the spontaneous magnetization of the leads. Some work, based on weaker fluctuation relations \cite{forster08}, has been published in Refs. \refcite{lopez12,lim13}. Ref. \refcite{spin-utsumi} attempted to apply stronger fluctuation relations to spintronics but overlooked the correct transformation law for spin currents under time reversal.

External magnetic fields and spontaneous magnetization are not the only ways how the time-reversal symmetry gets broken. It is  interesting to extend the results, discussed in this review, to systems with time-dependent Hamiltonians which are not invariant with respect to time reversal. Research in that direction includes Ref. \refcite{altland10}. Ref. \refcite{yi11} considers a system with a time-dependent magnetic field.
The ideas from Refs. \refcite{Maes1,Maes2} may be useful for the class of problems considered in this review.

Most research in the field has focused on quantum systems but there is nothing inherently quantum about time-reversal symmetry breaking. A discussion of fluctuation relations in classical systems with time-reversal symmetry breaking can be found in Refs. \refcite{class1,class2}.
An intriguing question involves possible applications of the results to biological systems with unidirectional transport.

\section*{Acknowledgments}

We thank M. Kardar and K. Saito for helpful discussions. C.W. acknowledges support from the NSF under Grant No. DMR-1254741. D.E.F. was supported in part  by the NSF under Grant No. DMR-1205715.

\end{document}